\documentclass[10pt,conference,a4paper]{IEEEtran}
\IEEEoverridecommandlockouts
\usepackage{cite}
\usepackage{amsmath,amssymb,amsfonts}
\usepackage{graphicx}
\usepackage{textcomp}
\usepackage{xcolor}
\usepackage{amssymb}
\usepackage{algorithm,algorithmicx,algpseudocode}
\usepackage{threeparttable}
\usepackage{multirow}
\usepackage{ae}
\graphicspath{{figs/}}

\makeatletter
\renewcommand{\maketag@@@}[1]{\hbox{\m@th\normalsize\normalfont#1}}%
\makeatother

\newtheorem{proposition}{Proposition}

\def\BibTeX{{\rm B\kern-.05em{\sc i\kern-.025em b}\kern-.08em
T\kern-.1667em\lower.7ex\hbox{E}\kern-.125emX}}

\begin{document}

\title{Marriage Matching-based Instant Parking Spot Sharing in Internet of Vehicles}

\author{
    \IEEEauthorblockN{
        Zhonghai Zhao\IEEEauthorrefmark{1}, {\em Student Member, IEEE}, Yang Xu\IEEEauthorrefmark{1}, {\em Member, IEEE}, Jia Liu\IEEEauthorrefmark{2}, {\em Member, IEEE}, \\
        Zhao Li\IEEEauthorrefmark{1}, {\em Member, IEEE}, Yulong Shen\IEEEauthorrefmark{1}, {\em Member, IEEE}, and Norio Shiratori\IEEEauthorrefmark{3}, {\em Life Fellow, IEEE}
    }
    \IEEEauthorblockA{
        \IEEEauthorrefmark{1}School of Computer Science and Technology, Xidian University, Xi'an, China \\
        \IEEEauthorrefmark{2}Center for Strategic Cyber Resilience Research and Development, National Institute of Informatics, Tokyo, Japan\\
        \IEEEauthorrefmark{3}Research and Development Initiative, Chuo University, Tokyo, Japan\\
        Email: zhzhaoXD@stu.xidian.edu.cn, yxu@xidian.edu.cn, jliu@nii.ac.jp, \\zli@xidian.edu.cn, ylshen@mail.xidian.edu.cn, norio@riec.tohoku.ac.jp}
}

\maketitle

\begin{abstract}
    The rapid development and integration of automotive manufacturing, sensor, and communication technologies have facilitated the emergence of the Internet of Vehicles (IoV). However, the explosive growing demand for parking spots has become a challenging issue to be addressed in IoV. In this paper, we propose a novel Smart Parking System (SPS) for IoV by applying the matching game approach. Specifically, the proposed SPS consists of three types of entities: drivers, parking spot owners (PSOs), and a certificate authority (CA). Drivers and PSOs send parking requests and parking spot information to the CA, respectively. The CA is responsible for identifying legitimate system users, matching drivers with PSOs, and recording their transaction information. We first design rational utility functions for drivers and PSOs, and then establish preference lists for them based on real-time conditions. With these information, we further develop an improved marriage matching (MM) algorithm to achieve stable matching results for drivers' requests and parking spot allocation. Simulation results demonstrate that the proposed MM-based SPS not only ensures stable matching results with the objective of distance minimization but also achieves an overall utility close to that of the optimal algorithm.
\end{abstract}


\section{Introduction}
The rapid development and integration of automotive manufacturing, sensor, and communication technologies have empowered the intelligence and interconnection of vehicles, leading to the emergence of Internet of Vehicles (IoV) as a new paradigm for intelligent transportation systems. However, in the context of urban development, the escalating disparity between the growing number of vehicles and the limited expansion of parking infrastructure has become a significant contributor to urban traffic congestion. Therefore, how to effectively manage urban resources to alleviate the scarcity of parking spots has become an imperative challenge that must to be addressed in IoV \cite{Zhu_TITS16}.

The parking space-sharing methodology holds great potential in improving the utilization of urban parking spots and alleviating traffic congestion in IoV. The fundamental idea is that parking spot owners (PSOs) have the opportunity to share their parking spots during idle periods to earn some profits, while drivers can achieve quick parking based on the instructions provided by the available leased parking space information \cite{Sun_TITS22}. The establishment of a parking information platform and the design of parking spot allocation and transaction mechanisms are crucial for implementing rational and efficient parking spot sharing. Consequently, they have received growing attention from both industry and academic communities.

By now, there have been some preliminary research works in the design of parking spot sharing systems \cite{Griggs_TITS15,Kim_TASE19,Huang_TV21,Wu_TITS22,Karaliopoulos_TITS23}. Specifically, Griggs \emph{et al.}\cite{Griggs_TITS15} designed a time allocation system for parking spots, taking into account customers' personal preferences during the matching process. Kim \emph{et al.}\cite{Kim_TASE19} proposed a distributed algorithm to minimize parking costs and balance multiple parking demands. Huang \emph{et al.}\cite{Huang_TV21} applied federated learning and reinforcement learning to parking spot selection, allowing for rapid sharing and prediction of parking spot information. Wu \emph{et al.}\cite{Wu_TITS22} presented an efficient parking spot search system based on mobile crowdsensing, enabling fast and efficient intelligent parking. More recently, Karaliopoulos \emph{et al.}\cite{Karaliopoulos_TITS23} developed a temporary parking spot leasing transaction scheme on an online parking reservation platform.

Although the aforementioned studies represent significant advancements in the design of shared parking systems, they either fail to consider the stability of the matching outcomes or overlook the heterogeneity of user preferences. However, in practical applications, stability is an important characteristic for evaluating the matching results, as it ensures the absence of cases where both parties have mutual preferences but are not matched. Moreover, addressing the heterogeneous requirements of different users (e.g., time conflict) is crucial for achieving the subdivision of parking resources and facilitating reasonable allocation, so as to improve social welfare.

Inspired by the above observations, in this paper, we aim to develop a novel smart parking system comprised of drivers, parking spot owners (PSOs), and a certificate authority (CA), leveraging matching game theory to facilitate the efficient sharing of available parking spots. We first design personalized utility functions for drivers and PSOs, and then construct preference lists taking into account the heterogeneity of both the customers' time requirements and the expected prices. Furthermore, we improve upon the traditional marriage matching algorithm to accommodate scenarios with incomplete preference lists, enabling stable matches between drivers and PSOs. The main contributions of this paper are summarized as follows:

\begin{itemize}
    \item
          We develop a smart parking system utilizing the principles of matching game theory. This system empowers drivers to accurately and reliably match with parking spots that best suit their specific needs and preferences, while enabling PSOs to profitably rent out their available parking spots through this system.

    \item
          We design personalized utility functions for drivers and PSOs, which can accurately characterize their interactions. Based on this, we create tailored preference lists for drivers and PSOs, allowing customers to dynamically adjust their preference vectors to achieve optimal parking spot matches based on their real-time conditions.

    \item
          We improve the traditional marriage matching (MM) algorithm and apply it to the driver-parking spot matching scenario. Simulation results show that the proposed MM-based smart parking system not only guarantees the stability of matching results but also achieves overall utility close to the optimal algorithm.
\end{itemize}

The remainder of this paper is organized as follows. Section~\ref{section:system_model} introduces the system overview and problem formulation. We elaborate on the marriage matching-based parking spot sharing algorithm design in Section~\ref{section:algorithm_design}. Section~\ref{section:simulation} presents the simulation results, followed by the conclusion in Section~\ref{section:conclusion}.

%-----------------------------------new section--------------------------------------%
\section{System Overview and Problem Formulation} \label{section:system_model}

\begin{figure}[!t]
    \centering
    \includegraphics[width=0.9\linewidth]{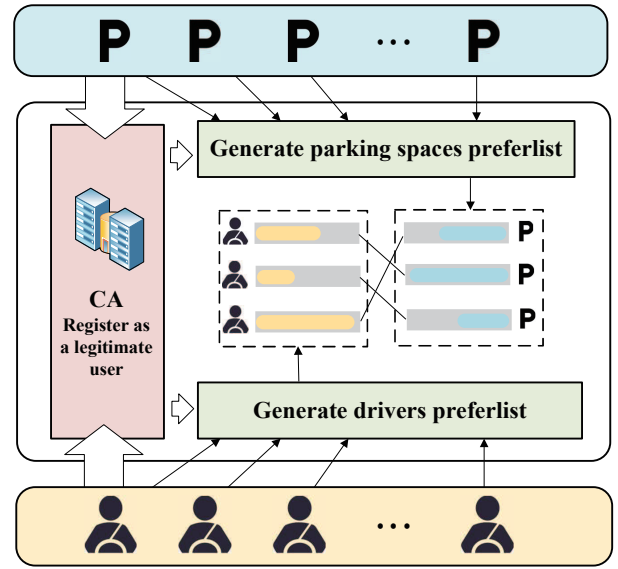}
    \caption{System Model.}
    \label{fig:system_model}
\end{figure}

We consider a smart parking system (SPS) in a vast urban area with multiple drivers and PSOs who can provide available parking spots. PSOs aim to earn profits during their free time by renting out their parking spots, while drivers seek to find available spots for parking their vehicles. In addition, we deploy a CA in the cloud to provide identity authentication and recognition services, as shown in Fig.~\ref{fig:system_model}.

\subsection{System Overview}

1) \emph{System Initialization}: The drivers and PSOs become legitimate entities after registering with CA in the SPS, and the CA initializes the basic information of drivers and PSOs. Let $D_i \in \mathcal{D}$ represent the $i$-th driver and $\mathcal{D}$ is the set of drivers. Let $P_j \in \mathcal{P}$ represent the $j$-th parking spot and $\mathcal{P}$ is the set of parking spots. The locations of driver $D_i$ and parking spot $P_j$ are denoted as $C_{D_i}$ and $C_{P_j}$, respectively. The expected parking time vector for driver $D_i$ is denoted as $\alpha_{D_i}=\left[a_1,a_2,\cdots,a_H\right]$, where each element represents a time interval of $24/H$ hours. For example, when $H = 48$, $a_1$ corresponds to the time interval 00:00-00:30, $a_2$ corresponds to the time interval 00:30$-$01:00, and so on. Each element of $\alpha_{D_i}$ takes value $0$ or $1$, i.e., when $a_h=1$, driver $D_i$ needs parking, and when $a_h=0$, there is no parking demand. Similarly, the sharable time of parking spot $P_j$ is denoted as a vector $\beta_{P_j}=\left[b_1,b_2,\cdots,b_H\right]$. If $b_h=1$, it indicates that the parking spot will not be rented out at this interval, and if $b_h=0$, it indicates that the parking spot is available at this interval.

2) \emph{Send Parking Request Message}: Once accomplishing the system initialization, a driver will send the following message to CA:
\begin{equation}
    Message_{D_i}=D_i||W_{D_i}(M_{\mathcal{P}})||W_{D_i}(R_{\mathcal{P}})||C_{D_i}||\alpha_{D_i},
\end{equation}
where $W_{D_i}(M_{\mathcal{P}})$ represents the maximum price per unit of time that driver $D_i$ is willing to pay for a parking spot in $\mathcal{P}$, and $W_{D_i}(R_{\mathcal{P}})$ represents the minimum reputation requirement that driver $D_i$ has for a parking spot in $\mathcal{P}$.

3) \emph{Send parking spot sharing message}: Each PSO submits the following message to CA:
\begin{equation}
    Message_{P_j}=P_j||M_{P_j}||W_{P_i}(R_{\mathcal{D}})||C_{P_j}||\beta_{P_j},
\end{equation}
where $M_{P_j}$ is the rental fee for a unit time specified by the PSO, and $W_{P_i}(R_{\mathcal{D}})$ refers to the minimum reputation requirement that parking spot $P_j$ imposes on a driver in $\mathcal{D}$.

4) \emph{Generate Preference Lists}: After both parties submit their information, the CA will generate corresponding preference lists for them according to their heterogeneous preferences.

5) \emph{Matching Result Output and Data Update}: The SPS applies a matching game approach (shown in the next section) to match the set of drivers with the set of parking spots, generating a matching result. Both parties conduct transactions according to the agreed-upon result. After the transaction is completed, they score each other, and the system updates their reputation values $R_{D_i}$ and $R_{P_i}$ based on the scores for the next round of matching reference. Specifically, the reputation value $R_{D_i}$ is updated by
\begin{equation}
    R_{D_i}=(1-\gamma )R'_{D_i}+\gamma \cdot r_{i,j}, \label{update}
\end{equation}
where $\gamma$ is a constant that satisfies $0 \leq \gamma \leq 1$, $r_{i,j}$ is the evaluation score assigned by $P_j$ to $D_i$ in the current transaction, and $R'_{D_i}$ is the last reputation value of $D_i$.
Similarly, the reputation value $R_{P_j}$ for a parking spot is updated by
\begin{equation}
    R_{P_j}=(1-\delta )R'_{P_j}+\delta \cdot r_{j,i}, \label{updatep}
\end{equation}
where $\delta \in (0,1)$ is a constant parameter, $r_{j,i}$ is the evaluation score assigned by $D_i$ to $P_j$ in the current transaction, and $R'_{P_j}$ is the last reputation value of $P_j$.

\subsection{Problem Formulation}
We consider that in IoV, drivers are more inclined to look for parking spots that are closer. Likewise, PSOs are also more inclined to rent out vacant parking spots to drivers who are closer in proximity (this helps to reduce the idle time of parking spots). We use $\mathcal{F}_{D_i, P_j}$ to denote the path length between the coordinate points of $D_i$ and $P_j$, i.e., $ \mathcal{F}_{D_i,P_j}=F(C_{D_i},C_{P_j})$, where the function $F(\cdot)$ calculates the distance between two coordinates. Therefore, the matching problem between drivers and PSOs in our smart parking system can be formulated as the following optimization problem:
\begin{equation}\label{yueshu}
    \begin{split}
        \min \,\, &\sum _{D_i \in \mathcal{D} ,P_j\in \mathcal{P}} \mathcal{F}_{D_i,P_j}.\\
        {\rm s.t.} \,\,& W_{D_i}(M_{\mathcal{P}}) \geqslant M_{P_j}, \forall_{D_i \in \mathcal{D}}, \forall _{P_j \in \mathcal{P}}\\
        &W_{D_i}(R_{\mathcal{P}}) \leqslant R_{P_j}, \forall_{D_i \in \mathcal{D}}, \forall _{P_j \in \mathcal{P}}\\
        &W_{P_i}(R_{\mathcal{D}}) \leqslant R_{D_i}, \forall_{D_i \in \mathcal{D}}, \forall _{P_j \in \mathcal{P}}\\
        &\alpha_{D_i}\cdot \beta_{P_j} = 0, \forall_{D_i \in \mathcal{D}}, \forall _{P_j \in \mathcal{P}}
    \end{split}
\end{equation}

%------------------------------------new section------------------------------------------%
\section{Marriage Matching-based Algorithm Design} \label{section:algorithm_design}

In this section, we first introduce the main concept of matching game theory. Then, we define the preference functions for drivers and PSOs and explain the details of creating preference lists for both entities in practice. Furthermore, we propose a stable driver-PSO matching scheme based on the improved marriage matching approach.

\subsection{Matching Game Concept}

\emph{Definition 1:} The outcome of the driver-PSO scheduling is a matching relation $\mu$, which is a function such that $\mathcal{D}\cup\mathcal{P}\rightarrow 2^{\mathcal{D}\cup\mathcal{P}} $ satisfies the following conditions:
\begin{itemize}
    \item $\mu(D_i) \subseteq \mathcal{P}$ such that $W_{D_i}(M_{\mathcal{P}}) \geqslant M_{P_j}$, $W_{D_i}(R_{\mathcal{P}}) \leqslant R_{P_j}$, and $\alpha_{D_i}\cdot \beta_{P_j} = 0$ for $\forall D_i \in \mathcal{D}$ and $\forall P_j \in \mathcal{P}$.
    \item $\mu(P_j) \subseteq \mathcal{D}$ such that $W_{P_i}(R_{\mathcal{D}}) \leqslant R_{D_i}$ and $\alpha_{D_i}\cdot \beta_{P_j} = 0$ for $\forall D_i \in \mathcal{D}$ and $\forall P_j \in \mathcal{P}$.
    \item $D_i \in \mu(P_j) \Longleftrightarrow \mu(D_i) = P_j$ for $\forall D_i \in \mathcal{D}$ and $\forall P_j \in \mathcal{P}$.
\end{itemize}

\emph{Definition 2:} Let $\mathcal{M}$ denote a matching result in SPS, which is a subset of $\mu$. We say that $D_1$ is assigned to $P_2$ or $P_2$ is assigned to $D_2$ if the pair $(D_1,P_2) \in \mathcal{M}$. Let $\mathcal{H} = \mathcal{D} \times \mathcal{P} $, a pair $(D_i,P_j) \in \mathcal{H} \setminus  \mathcal{M} $ blocks $\mathcal{M}$, or $(D_i,P_j)$ is a blocking pair for $\mathcal{M}$, if the following conditions are satisfied:
\begin{itemize}
    \item $D_i$ is unassigned or prefers $P_j$ to $\mathcal{M}(D_i)$.
    \item $P_j$ is unassigned or prefers $D_i$ to $\mathcal{M}(P_j)$.
\end{itemize}

\emph{Definition 3:} The matching relation $\mu$ is considered stable if (1) there are no blocking pairs, and (2) except for a few unmatched drivers, the majority of drivers in set $\mathcal{D}$ are assigned to parking spaces in set $\mathcal{P}$.

\subsection{Preference Functions of Drivers and PSOs}
In the smart parking system, we consider that drivers in need of parking have a preference for the nearest available parking spot, which allows them to minimize both the time and fuel costs associated with finding a suitable parking spot.  Therefore, for each driver $D_i \in \mathcal{D}$, there is a strict transitive preference relation $L_{D_i}(\mathcal{P})$ on the set of $\mathcal{P}$ regarding parking spots. The preference relation $P_j \succ_{D_i} P_k $ means that driver $D_i$ prefers parking spot $P_j$ over parking spot $P_k$. Based on the above definition, we can define the driver's preference function as follows:
\begin{equation}
    P_j \succ_{D_i} P_k \Leftrightarrow L_{D_i}(P_j) < L_{D_i}(P_k),
\end{equation}
where
\begin{equation}\label{disdp}
    L_{D_i}(P_j) = \mathcal{F}_{D_i,P_j}.
\end{equation}

The preference list for each parking spots $P_j \in \mathcal{P}$ is also based on distance because the shorter the distance, the faster the parking spot can execute transactions, reducing idle time and increasing profits. Similarly, the preference function for parking spots can be expressed as follows:
\begin{equation}
    D_i \succ_{P_j} D_k \Leftrightarrow L_{P_j}(D_i) < L_{P_j}(D_k),
\end{equation}
where
\begin{equation}\label{dispd}
    L_{P_j}(D_i) = \mathcal{F}_{P_j,D_i}.
\end{equation}

\subsection{Preference List Generation}

We summarized in Algorithm~\ref{algorithm:Driver_Preference.} how to create a preference list for each driver based on the real-time system conditions. The algorithm takes all registered parking spots in the city as input and outputs a preference list $L(D_i)$ for each driver $D_i$. Algorithm~\ref{algorithm:Driver_Preference.} first iterates through each parking spot (line 1) and then checks whether the parking spot and driver $D_i$ satisfy the constraints in the optimization problem (\ref{yueshu}) (line 3). Parking spaces that meet the constraints are added to the list $L(D_i)$. Finally, the $L(D_i)$ list is sorted in ascending order according to expression (\ref{disdp}) (line 8).

\begin{algorithm}[!ht]
    \caption{Driver Preference List Generation.}
    \label{algorithm:Driver_Preference.}
    \begin{algorithmic}[1]
        \Require
        Registered parking spot set $\mathcal{P}$
        \Ensure
        Preference list $L(D_i)$ of driver $D_i$;
        \While {there are non-visited parking spots in $\mathcal{P}$}
        \State {Select a non-visited $P_j$ and mark it as visited};
        \If {$D_i$ and $P_j$ meet the constraints specified in (\ref{yueshu})}
        \State {Record the distance $\mathcal{F}_{D_i,P_j}$ between $D_i$ and $P_j$};
        \EndIf
        \EndWhile
        \State {Use expression~(\ref{disdp}) to rank the parking spots and store the results in $L(D_i)$}; \\
        \Return $L(D_i)$;
    \end{algorithmic}
\end{algorithm}

Similarly, we can generate a preference list $L(P_j)$ for each PSO, as summarized in Algorithm~\ref{algorithm:PSO_Preference.}.

\begin{algorithm}[!ht]
    \caption{PSO Preference List Generation.}
    \label{algorithm:PSO_Preference.}
    \begin{algorithmic}[1]
        \Require
        Registered driver set $\mathcal{D}$
        \Ensure
        Preference list $L(P_j)$ of parking spot $P_j$;
        \While {there are non-visited drivers in $\mathcal{D}$}
        \State {Select a non-visited driver $D_i$ and mark it as visited};
        \If {$P_j$ and $D_i$ meet the constraints specified in (\ref{yueshu})}
        \State {Record the distance $\mathcal{F}_{P_j,D_i}$ between $P_j$ and $D_i$};
        \EndIf
        \EndWhile
        \State {Use expression~(\ref{dispd}) to rank the drivers and store the results in $L(P_j)$}; \\
        \Return $L(P_j)$;
    \end{algorithmic}
\end{algorithm}

\subsection{Improved Marriage Matching Algorithm}

Note that from the process of constructing the preference list, it can be observed that the generated preference list is incomplete. In other words, for different drivers, the set of parking spots that can be matched may vary, and vice versa. To address this issue, we will design a solution based on the Gale-Shapley stable marriage matching algorithm~\cite{gale2013college}, which can accommodate incomplete preference lists.

We consider the driver as the party initiating the request and prioritize the driver's interest. Therefore, we iterate through the set of drivers $\mathcal{D}$ and send matching requests to parking spots in order of decreasing preference within each driver's preference list.

\begin{figure}[htbp]
    \centerline{\includegraphics[scale=0.4]{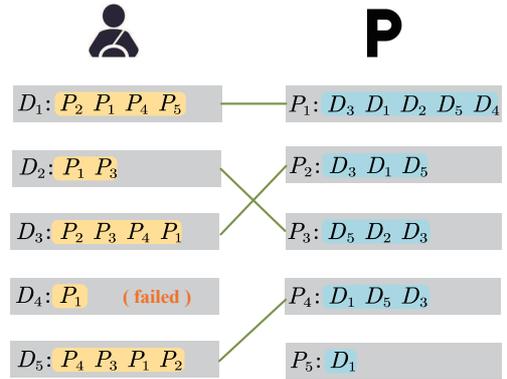}}
    \caption{Matching algorithm model.}
    \label{fig:matching_algorithm_model}
\end{figure}

In Fig~\ref{fig:matching_algorithm_model}, we observe that without loss of generality, $D_1$ is chosen as the first driver. According to $D_1$'s preference list, a request is made for the second parking spot $P_2$. Since $P_2$ has not been matched yet, a temporary pairing is made between $D_1$ and $P_2$. Next, when driver $D_2$ is selected, he sends a request to the parking lot set based on his preference list. Similar to the request made by $D_1$, $D_2$ is temporarily paired with the first parking spot $P_1$. When $D_3$ is selected, it is found that the first parking spot $P_2$ in $D_3$'s preference list has already been assigned to $D_1$. However, in $P_2$'s preference list, $D_3$ has a higher priority than $D_1$. According to Definition 2, $(D_3, P_2)$ forms a blocking pair. Therefore, to eliminate the blocking pair, $D_1$ is unpaired from $P_2$ (and will continue requesting the next parking spot $P_1$ in the next round), while $P_2$ is temporarily paired with $D_3$. Similarly, in $D_4$'s turn, it is found that $D_2$ and $P_1$ have already been paired. Upon checking $P_1$'s preference list, it is discovered that $D_2$ has a higher priority than $D_4$, and thus $D_4$'s request fails, and he continues to request a parking spot in the next round. If a driver's preference list has been fully traversed and no parking spot has been found for a pairing, it is considered a failure. Algorithm 3 provides specific implementation details. The variables $\left|\mathcal{D}\right|$ and $\left|\mathcal{P}\right|$ represent the number of drivers and the number of parking spots in the same area, respectively.

\begin{algorithm}[!ht]
    \caption{PSO Preference List Generation.}
    \label{algorithm:MM}
    \begin{algorithmic}[1]
        \Require
        $\mathcal{D}$, $\mathcal{P}$
        \Ensure
        matching result set $\mathcal{M}$
        \State{Initial Phase: An empty circular queue $que$ is initialized with all elements of set $\mathcal{D}$, matching result set $\mathcal{M}$ initialized as empty, the variable $n$ is initialized to 0;}
        \For{ i = 1:$|\mathcal{D}|$ }
        \State Build the preference list $L(D_i)$ through algorithm 1
        \EndFor
        \For{ j = 1:$|\mathcal{P}|$ }
        \State Build the preference list $L(P_j)$ through algorithm 2
        \EndFor

        \While {$n \neq$ $|\mathcal{D}|$}
        \State Pop a driver from $que$ as $D_i$;
        \State Look the next parking lot $P_j$ in the $D_i$'s preference list;
        \If {$\exists P_j$}
        \If {$\exists\mathcal{M}(P_j)$}
        \If{ $D_i$ is more preferences than the old matched driver $\mathcal{M}(P_j)$ in $P_j$'s preference list $L_(P_j)$ }
        \State Remove the old pair $(\mathcal{M}(P_j),P_j)$ from $\mathcal{M}$, a new pair $(D_i,P_j)$ add to $\mathcal{M}$. The old driver is pushed into the $que$;
        \Else
        \State Request failed, $D_i$ is pushed into the $que$
        \EndIf
        \Else
        \State A new pair $(D_i,P_j)$ add to $\mathcal{M}$
        \State $n++$
        \EndIf
        \Else
        \State $n++$
        \EndIf
        \EndWhile \\
        \Return $\mathcal{M}$;
    \end{algorithmic}
\end{algorithm}

After several rounds of matching, we will get the final matching result for the specific example above $(D_1, P_1), (D_2, P_3), (D_3, P_2), (D_5, P_4)$, and $D_4$ fails to match.

\begin{proposition}
    For algorithm 3, the proposed method based on a stable marriage matching algorithm can converge and obtain stable matching results. Besides, as for the performance, the running time of the implementation can achieve $\mathcal{O}(\left|\mathcal{D}\right|\times \left|\mathcal{P}\right|)$, where $(\left|\mathcal{D}\right|\times \left|\mathcal{P}\right|)$ denotes all the possible matching pairs.
\end{proposition}

\begin{IEEEproof}
    Let $\mathcal{M}$ be an instance of SPS matching. The stable pairs in $\mathcal{M}$ can be found in $\mathcal{O}(\left|\mathcal{D}\right|)$ time. The stable matchings in $\mathcal{M}$ can be listed in $\mathcal{O}(\left|\mathcal{P}\right|)$ time per matching, after $\mathcal{O}(\left|\mathcal{D}\right|)$ preprocessing time. Therefore, the overall running time can achieve $\mathcal{O}(\left|\mathcal{D}\right|\times \left|\mathcal{P}\right|)$. The corresponding proof can be found in \cite{gent2002empirical} and \cite{4460480} in detail.
\end{IEEEproof}

\section{Simulation Results} \label{section:simulation}

In this section, we evaluated the efficiency of our proposed matching algorithm. We compared its performance with that of greedy matching and random matching and also compared our proposed matching method with the theoretically optimal algorithm.

Random matching means that for a selected driver, a parking spot is randomly selected from his preference list for matching. If the parking spot is unmatched, the driver is matched with it; if the parking spot is matched and is greater than the maximum random count then the driver fails the match and continues to select the next driver, otherwise, the parking spot is re-randomized and matched again. Since preference lists vary in length and the matching result depends on the order of the selected drivers, to have as many pairs of matches as possible, we first sort the list of drivers in ascending order of preference lists and allow the maximum number of randoms for the same driver to equal its list length.

Greedy matching algorithm is based on the random matching algorithm sorting, front to back to select the driver, for this driver and then priority from the preference list of the first to take out the parking lot, if the parking lot has not yet matched, the driver and parking spot to match, if the parking spot has been matched, the driver give up this parking spot, continue to query the second parking spot in the preference list, cycle and repeat, if the final still can not find a parking spot, then the driver match failed.

To achieve the theoretical maximum utility, we use the Hungarian algorithm\cite{kuhn1955hungarian}, which aims to find the sum of the weights on the edges that maximize while making as many points match as possible in a bipartite graph with weighted edges. In this article, we need to process distances. We subtract the actual distance from the distance upper bound and then perform the optimal bipartite matching on the result, and the resulting matching is the one that minimizes the total distance.

We randomly generate the distance $\mathcal{F}_{D_i,P_j}$ between parking spots and drivers between $\left[0,5\right]$ km, and the edges between drivers and parking spots are randomly generated based on the percentage of their relationship $\eta _\mathcal{E} = \frac{\left|\mathcal{E}\right|}{\left|\mathcal{E}_{max}\right|}$, where $\left|\mathcal{E}\right|$ is the number of actual relationships in a match and the maximum theoretical number of relationships $\left|\mathcal{E}_{max}\right|$ is equal to $\left|\mathcal{D}\right|\cdot\left|\mathcal{P}\right|$. The edge weight is the distance between the driver and the parking spot. Based on this, we generate preference lists for both drivers and parking spots.

\begin{figure}[t]
    \centering
    \includegraphics[width=0.45\textwidth]{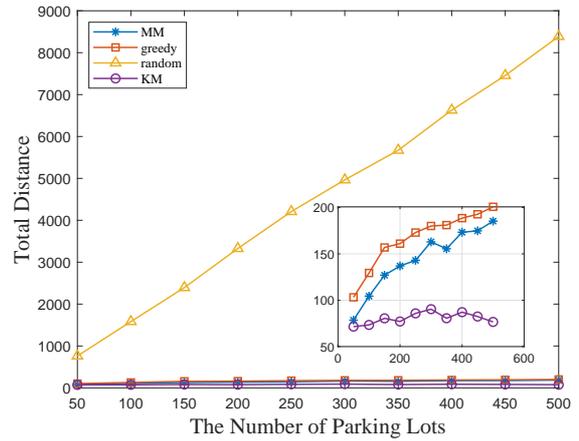}
    \caption{Total distance with parking spots' number.}
    \label{fig:Utility_parking spot}
\end{figure}

First, we discuss the impact of increasing the data scale on the total distance under the condition of an equal number of drivers and parking spots. Here we set the $\eta _\mathcal{E}=20\%$, and the results are shown in Fig. \ref{fig:Utility_parking spot}. We enumerate cases where the number of drivers or parking spots increased from 50 to 500. As the data size increased, the random algorithm's total distance grew much faster than that of the greedy algorithm, MM algorithm, and KM algorithm. The next fastest growth rate was the greedy algorithm. Our algorithm's final total distance was second only to the optimal algorithm.

\begin{figure}[t]
    \centering
    \includegraphics[width=0.45\textwidth]{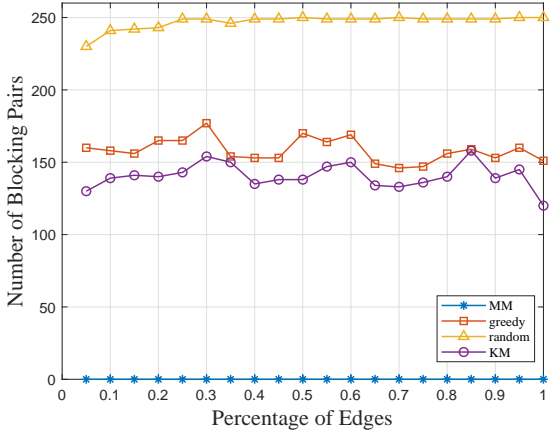}
    \caption{Number of blocking pairs.}
    \label{fig:blocking_pairs}
\end{figure}

In Fig. \ref{fig:blocking_pairs}, we set the number of drivers and parking spots to be 250 each and obtained trend lines for the number of blocking pairs as the $\eta _\mathcal{E}$ increases from 5\% to 100\% using four different algorithms.

In MM algorithm, it can be seen that the number of blocking pairs is always zero, while in the KM algorithm, there are blocking pairs between 100 and 150. It can be seen that the optimal matching algorithm for bipartite graphs is not stable. However, satisfying the matching between drivers and parking spots is an important means to improve social well-being. Therefore, the MM algorithm makes certain sacrifices in achieving matching stability to minimize distance.

\begin{figure}[t]
    \centering
    \includegraphics[width=0.45\textwidth]{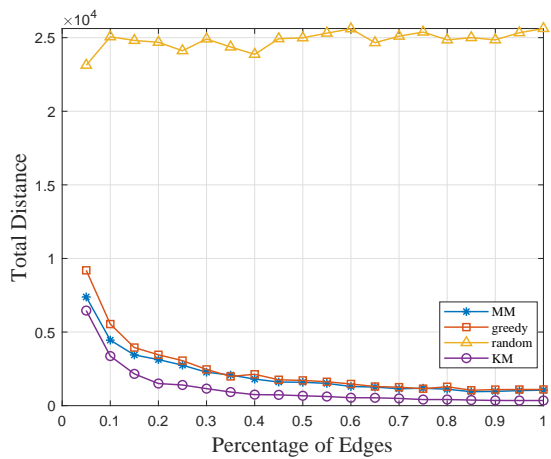}
    \caption{Total distance with edges' percentage.}
    \label{fig:Utility_percentage}
\end{figure}

During the simulation process, we found that the matching algorithm in this paper is less sensitive to the $\eta _\mathcal{E}$. Therefore, by designing experiment Figure \ref{fig:Utility_percentage}, we discuss the trend of the total distance of four matching algorithms as the $\eta _\mathcal{E}$ increases from 5\% to 100\% when the number of drivers is 250, the ratio of drivers to parking spots remains at $1:1$, and the distance between drivers and parking spots is initialized between [0,100] km.

From the Fig. \ref{fig:Utility_percentage}, it can be seen that the algorithm in this paper and the greedy algorithm are both very close to the optimal algorithm in terms of matching results. The random matching algorithm performs the worst, with a total distance much larger than the other algorithms. By observing the distance curve of the algorithm in this paper, it can be found that when the $\eta _\mathcal{E}$ is higher than 20\%, the total distance of the matching result tends to be horizontal but decreases slowly. This is because when there are more contacts, there are more matching objects for both drivers and parking spots with shorter distances. However, for the part below 20\%, the total distance will increase but still be smaller than that of the random algorithm. Therefore, even when there are relatively few available parking spots (low relationship percentage), this algorithm can still achieve a relatively low overall total distance.

\begin{table}[htbp]
    \caption{Algorithm Running Time Comparison}
    \begin{center}
        \begin{tabular}{|c|c|c|c|c|c|}
            \hline
            \multirow{2}{*}{\textbf{Algorithm}} & \multicolumn{5}{|c|}{\textbf{Data Scale}}                                                                                                 \\
            \cline{2-6}
                                                & \textbf{\textit{100}}                     & \textbf{\textit{200}} & \textbf{\textit{300}} & \textbf{\textit{400}} & \textbf{\textit{500}} \\
            \hline
            random                              & 0.0012                                    & 0.0026                & 0.0037                & 0.0053                & 0.0070                \\
            \hline
            greedy                              & 0.0016                                    & 0.0040                & 0.0050                & 0.0072                & 0.0094                \\
            \hline
            MM                                  & 0.0061                                    & 0.0209                & 0.0460                & 0.0908                & 0.1643                \\
            \hline
            KM                                  & 1.0984                                    & 1.6678                & 11.2773               & 16.5195               & 44.1794               \\
            \hline

            \multicolumn{4}{l}{The unit of time is second.}
        \end{tabular}
        \label{tab1}
    \end{center}
\end{table}

In TABLE \ref{tab1}, we simulated the running time of four algorithms. We set the number of drivers from 100 to 500, with the same number of parking spots as drivers and $\eta _\mathcal{E}$ set to 20\%. The data in Table 1 was obtained accordingly. It is worth noting that when the number of drivers reaches 500 (the total number of parking spots and drivers is 1000), the MM algorithm matching time reaches about 0.2s. However, the KM algorithm, which can achieve the shortest distance, requires approximately 44.2s to complete the matching. In a system that requires real-time matching, the system's response time directly affects user experience. Although the KM algorithm can obtain the optimal matching result, it spends a lot of time, which is not worthwhile.

\section{Conclusion}  \label{section:conclusion}

In this paper, we design SPS to provide an idea for the rational allocation of shared parking spot resources. First, we analyzed the urgent demand for resource sharing in the current social environment and proposed a smart parking system model, then designed the utility functions of drivers and parking spots according to the actual situation. And then optimized our driver parking spot matching algorithm based on the original marriage matching algorithm. Finally, the performance of the modified algorithm is verified by experimental simulation.

\bibliographystyle{IEEEtran}
\bibliography{IEEEabrv,Globecom}

\end{document}